\documentclass[twocolumns,bibyear]{aa} 
\usepackage{graphicx}
\usepackage{txfonts}
\def\Q{\ifmmode\mathcal{Q}\else$\mathcal{Q}$\fi}

\begin{document} 
   \title{Star formation activity and the spatial distribution and mass segregation of dense cores in the early phases of star formation}\titlerunning{Structure and mass segregation in the early phases of star formation}

   \author{Sami Dib \inst{1,2}, Thomas Henning\inst{2}}

   \institute{Laboratoire d'Astrophysique de Bordeaux, Universit\'{e} de Bordeaux, CNRS, B18N,  all\'{e}e Geoffroy Saint-Hilaire, 33615, Pessac, France
     \and
                 Max Planck Institute for Astronomy, K\"{o}nigstuhl 17, D-69117, Heidelberg, Germany\\
                      \email{sami.dib@gmail.com}                  
          }

 
\abstract{We examine the spatial distribution and mass segregation of dense molecular cloud cores in a number of nearby star forming regions (the region L1495 in Taurus, Aquila, Corona Australis, and W43) that span about four orders of magnitude in star formation activity. We use an approach based on the calculation of the minimum spanning tree, and for each region, we calculate the structure parameter \Q\ and the mass segregation ratio $\Lambda_{\rm MSR}$ measured for various numbers of the most massive cores. Our results indicate that the distribution of dense cores in young star forming regions is very substructured and that it is very likely that this substructure will be imprinted onto the nascent clusters that will emerge out of these clouds. With the exception of Taurus in which there is nearly no mass segregation, we observe mild-to-significant levels of mass segregation for the ensemble of the 6, 10, and 14 most massive cores in Aquila, Corona Australis, and W43, respectively. Our results suggest that the clouds' star formation activity are linked to their structure, as traced by their population of dense cores. We also find that the fraction of massive cores that are the most mass segregated in each region correlates with the surface density of star formation in the clouds. The Taurus region with low star-forming activity is associated with a highly hierarchical spatial distribution of the cores (low \Q\ value) and the cores show no sign of being mass segregated. On the other extreme, the mini-starburst region W43-MM1 has a higher \Q\ that is suggestive of a more centrally condensed structure and it possesses a higher fraction of massive cores that are segregated by mass. While some limited evolutionary effects might be present, we largely attribute the correlation between the star formation activity of the clouds and their structure to a dependence on the physical conditions that have been imprinted on them by the large scale environment at the time they started to assemble.}
 

 \maketitle
%

\section{Introduction}

The star formation process yields a large number of physical quantities and distribution functions that can be quantified by observations. Comparing these quantities and distributions between different star forming regions and/or with theoretical models and numerical simulations helps us gain insight into the relevance of various physical processes in different galactic environments. Some of the most studied quantities are the mass distribution of dense cores and of the stellar initial mass function in the early and emerging phases of the formation of stellar clusters (e.g., Johnstone \& Bally 2006; Dib 2014; Hony et al. 2015; Dib et al. 2017). The spatial distribution of dense cores in the early phases of star formation, of protostars in the phases of a cluster's buildup, and of stars in the (pseudo)gas-free phase of a young cluster, in conjunction with their masses and dynamics can also encapsulate critical information on how clusters assemble and form in different environments. 

The spatial distribution of stars in young and in evolved clusters has received a substantial amount of attention both in observations (McNamara \& Sekiguchi 1986; S\'{a}nchez \& Alfaro 2009; Gouliermis et al. 2014; Parker \& Alves de Oliveira 2017; Dib et al. 2018) and in numerical simulations of star forming regions (Schmeja \& Klessen 2006; Lomax et al. 2011; Parker \& Dale 2015; Gavagnin et al. 2017). However, quantifying the structure of star forming regions in the very early phases of star formation has remained elusive. This was principally due to the scarcity of observational data with the adequate spatial resolution to probe core masses in the stellar mass regime. With the advent of the {\it Hershel space observatory} (hereafter Herschel) and the {\it Atacama Large Millimeter Array} of radiotelescopes (ALMA), it is now possible to probe the spatial and mass distributions of dense structures in nearby star forming regions down to the mass regime of proto-brown dwarfs. The high sensitivity and spatial resolution of both Herschel and ALMA has provided so far unprecedented insight into quantities such as the dense core mass function (CMF) in the regions of Aquila, Taurus, Corona Australis, and W43 (Andr\'{e} et al. 2010; K\"{o}nyves et al. 2010,2015; Marsh et al. 2016; Bresnahan et al. 2018; Motte et al. 2018). The published CMFs of these regions suggest, at the very least, a striking difference between the low mass star forming regions such as Aquila and Taurus, and regions of massive star formation such as W43. When described by a power law of the form $dN/dM \propto M^{-\alpha}$, the derived value of $\alpha$ for W43 is $\approx 1.9$ which makes the CMF of W43 ostensibly shallower than in low mass star forming regions for which the derived values of $\alpha$ close to $2.3$ (Motte et al. 1998, but see Sadavoy et al. 2010), or even steeper as in the case of the California Molecular Cloud (Zhang et al. 2018). Dib et al. (2008a) showed that the slope of the CMF is steeper for cores that are defined using molecular species that trace higher densities of the gas and that are associated with an increasing degree of gravitational boundedness of the cores. However, in the case of the aforementioned star forming regions, the differences in the slopes of the CMF cannot be attributed to differences in the choice of the density tracers as all of the cores in those regions are observed in the continuum submillimeter emission. Within individual star forming regions, it has now been established that there are environmental differences between the populations of cores that are found on and off the filamentary structure of the clouds (Polychroni et al. 2013; Olmi et al. 2016; Kainulainen et al. 2017; Bresnahan et al. 2018). This can be attributed to effects of early mass segregation and/or to an extended phase of gas accretion by cores that reside inside the filaments (Dib et al. 2010) . 
 
\begin{figure*}
\centering
\includegraphics[width=0.46\textwidth]{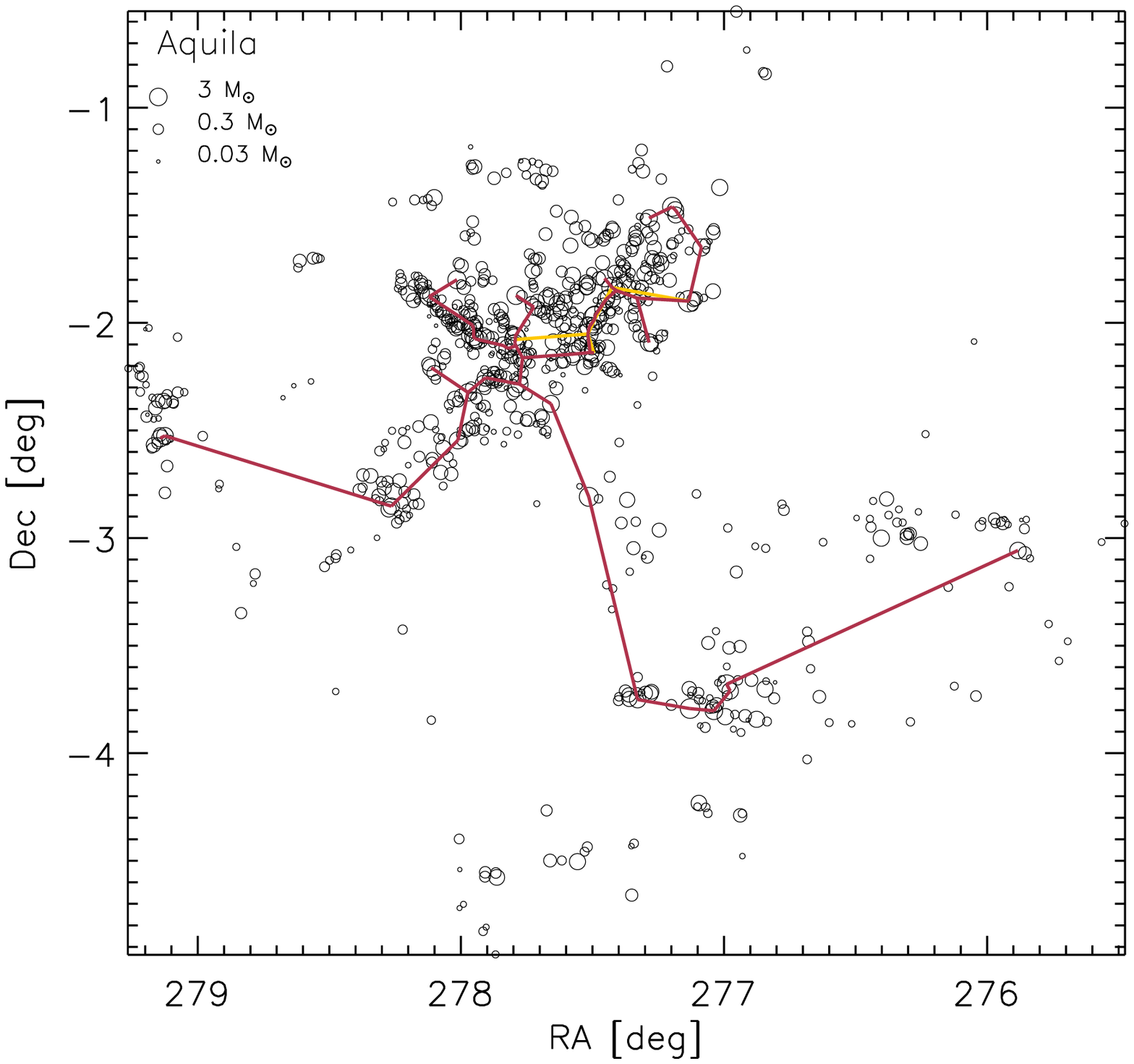}
\includegraphics[width=0.46\textwidth]{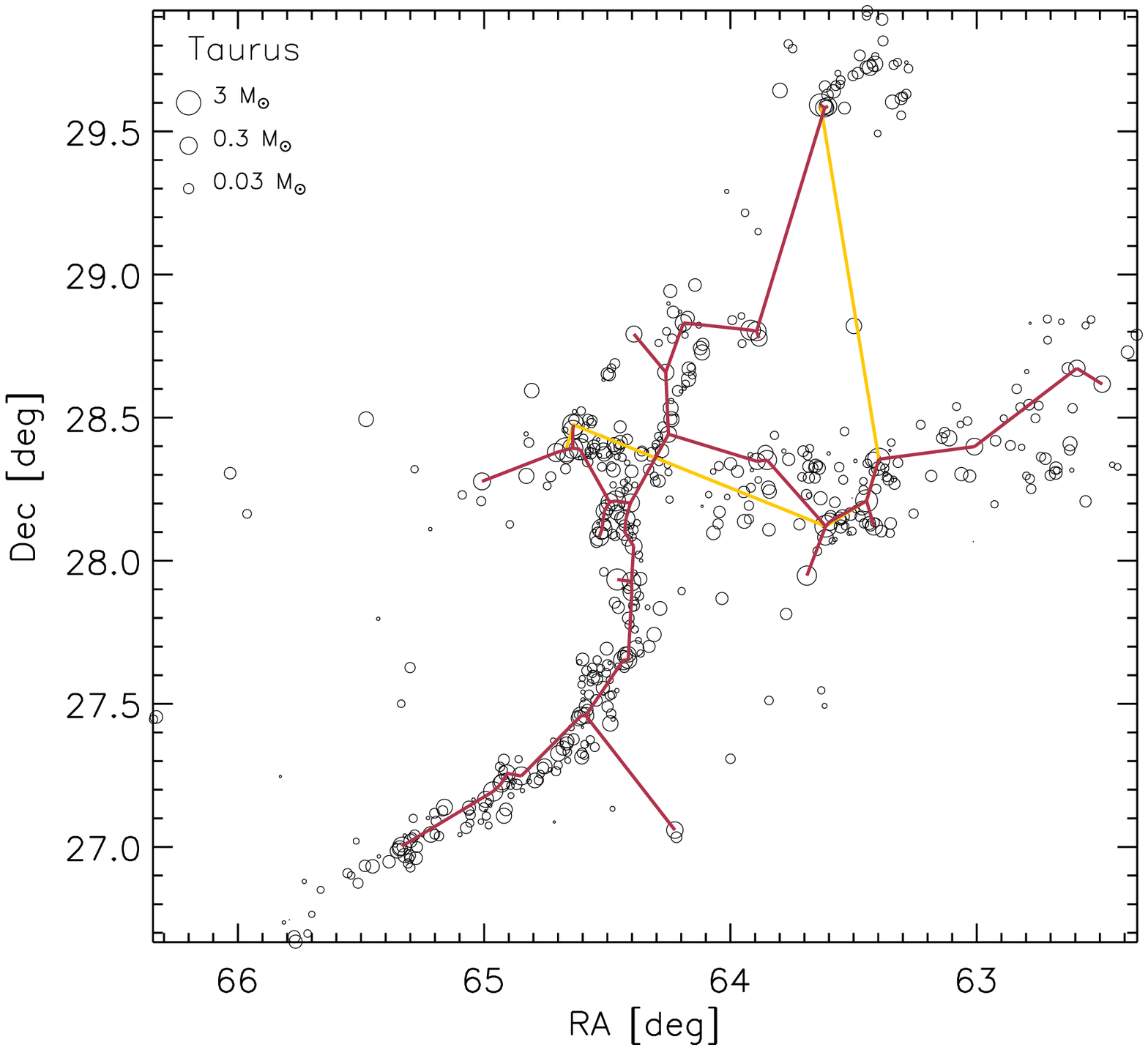}\\
\includegraphics[width=0.46\textwidth]{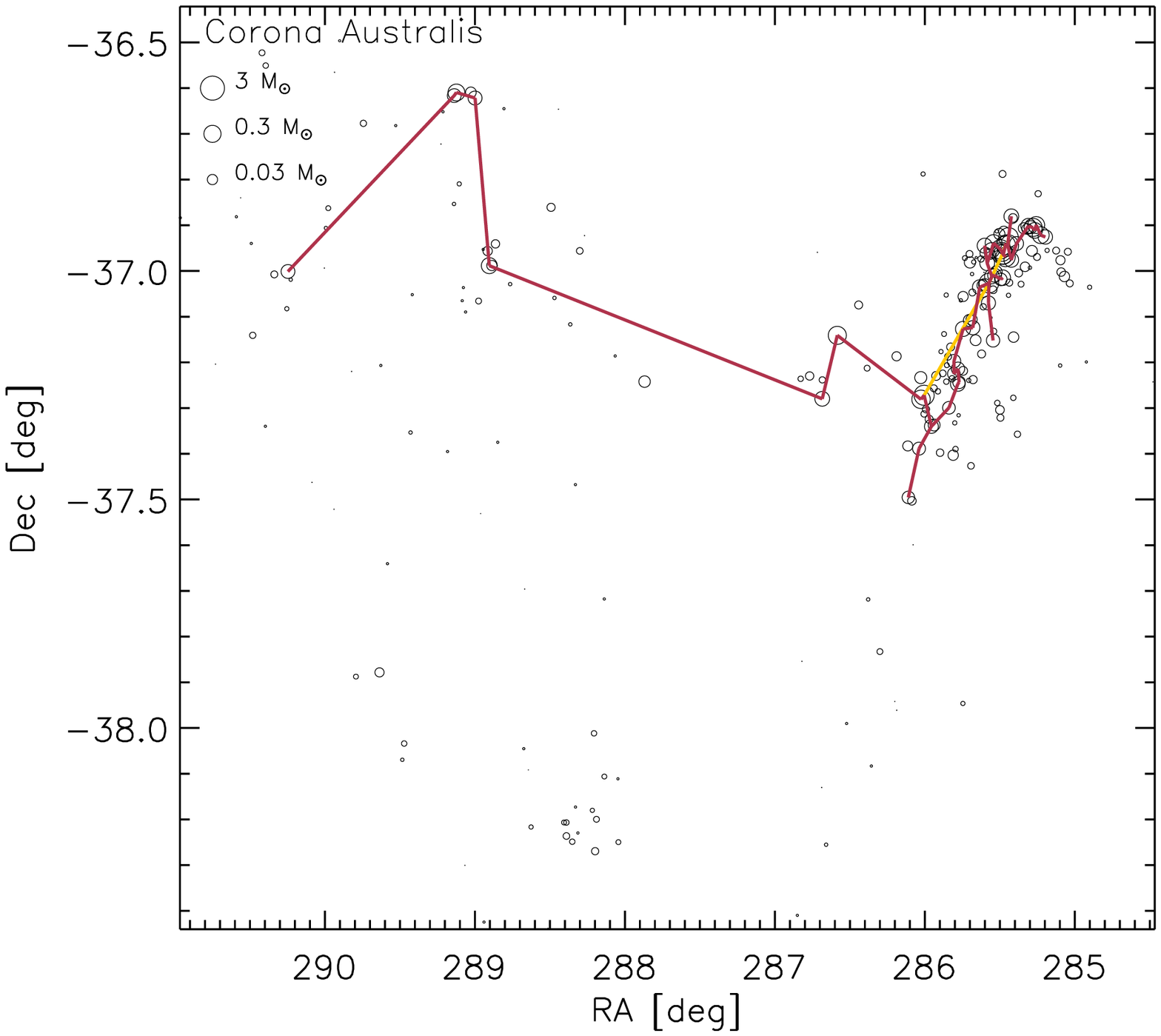}
\includegraphics[width=0.46\textwidth]{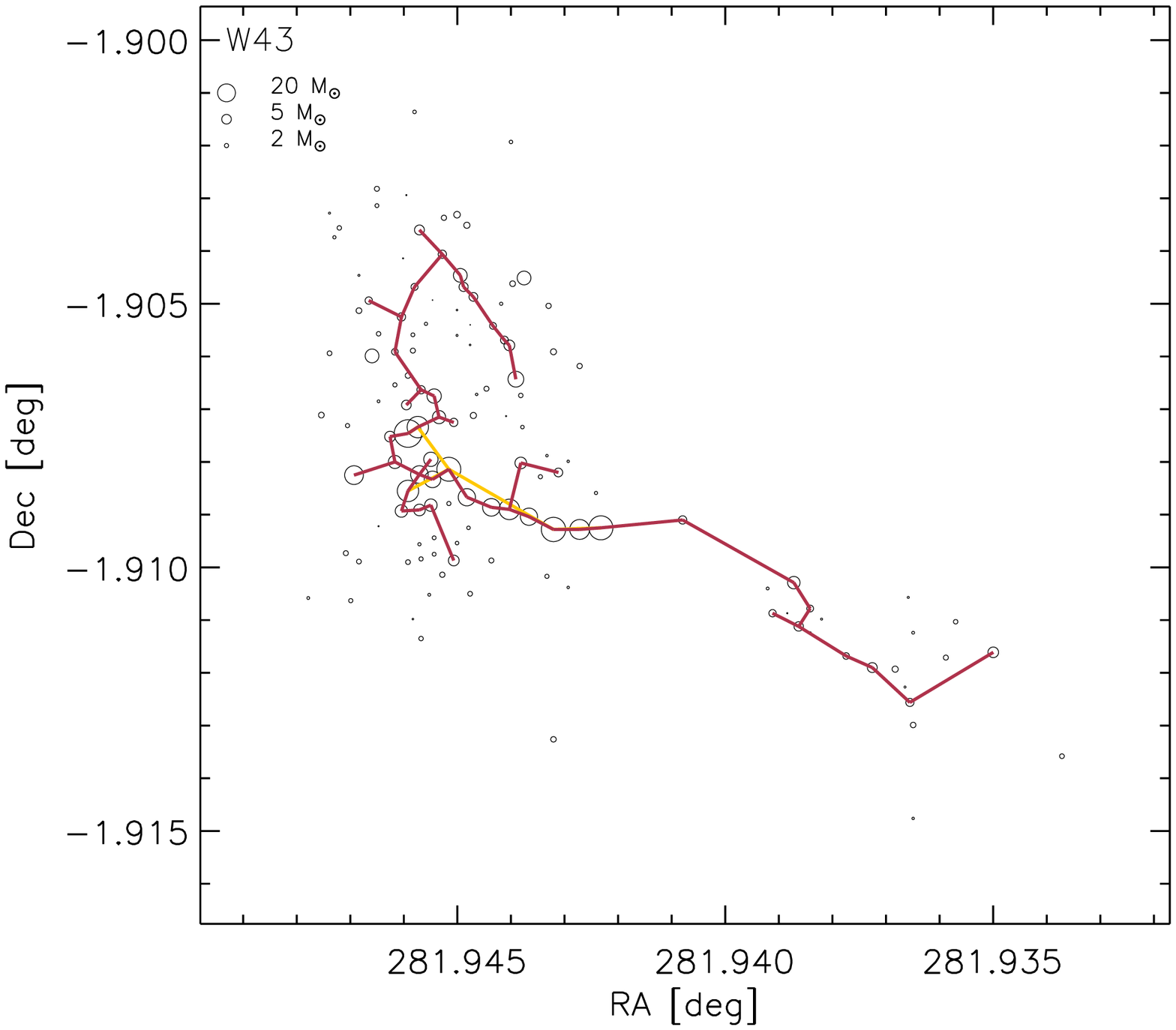}
\caption{Spatial distribution of dense cores in the four star forming regions considered in this work. The sizes of cores have been scaled (with an arbitrary formula) according to their masses (i.e., larger sizes relate to more massive cores) in order to visually highlight the location of the most massive cores.The minimum spanning tree (MST) for the ensemble of the 6 and 50 most massive cores in each region are displayed with the yellow and purple lines, respectively.}
\label{fig1}
\end{figure*}
   
\begin{figure}
\centering
 \includegraphics[width=0.9\columnwidth]{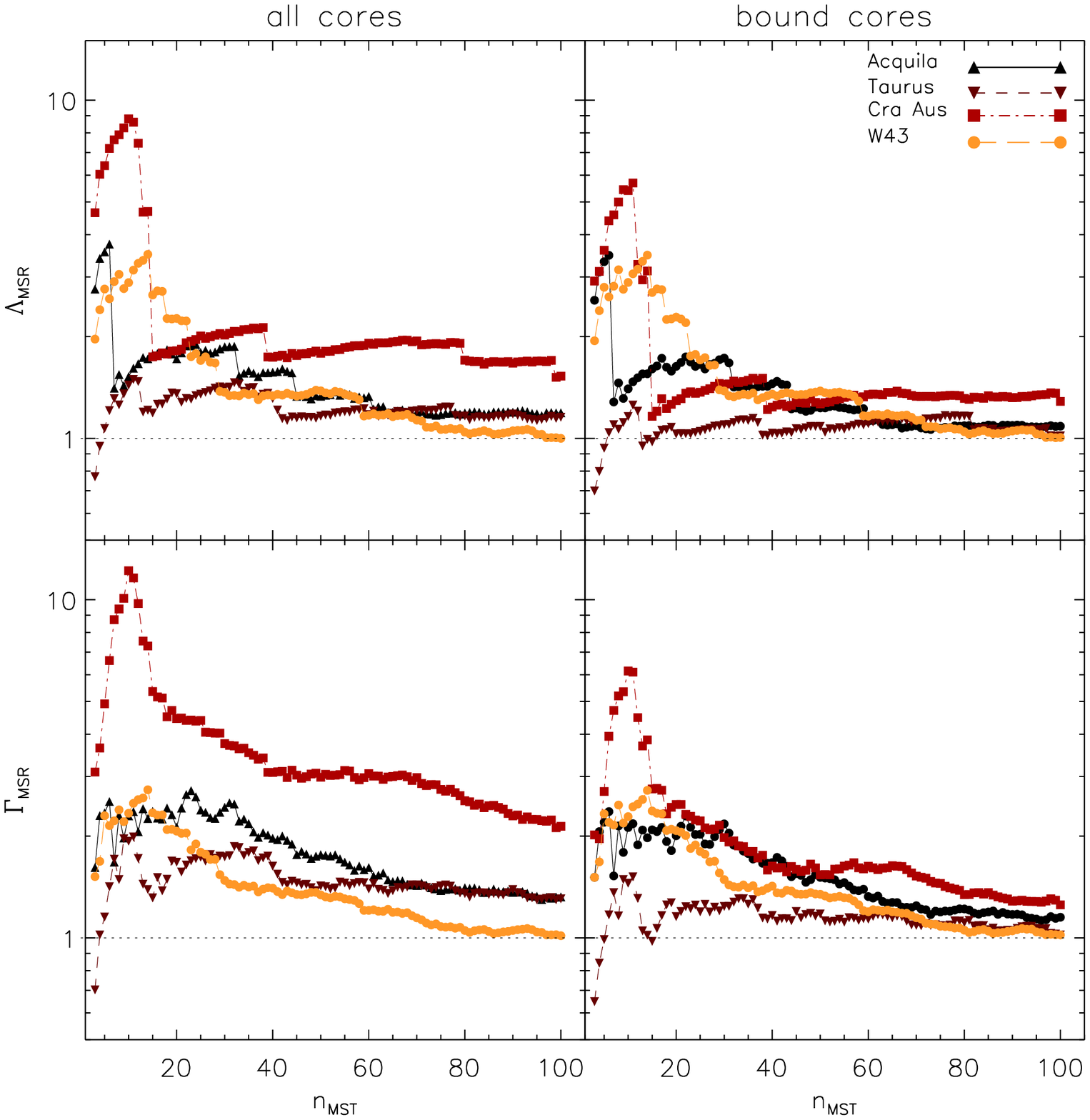}
 \vspace{0.5cm}
 \caption{The mass segregation ratios $\Lambda_{\rm MSR}$ (Allison et al. 2009, top row) and $\Gamma_{\rm MSR}$ (Olczak et al. 2011, bottom row) as a function of the number of most massive cores used in computing them, $n_{\rm MST}$. The values of $\Lambda_{\rm MSR}$ and $\Gamma_{\rm MSR}$ are calculated for the entire population of cores in the star forming regions (left column) and for the populations of bound cores (prestellar cores and cores with a protostar). A version of this figure that includes estimates of the uncertainties on $\Lambda_{\rm MSR}$ and $\Gamma_{\rm MSR}$  for each individual region is shown in Fig.~\ref{figappb1}.}
 \label{fig2}
\end{figure}

We complement the picture provided by the CMF of these different regions by analyzing the spatial distribution of their populations of dense cores. The structure and mass segregation of the cores in these regions lock important information on how star clusters form and evolve in different galactic environments. A number of studies have suggested a similarity between the structure of young clusters and of molecular clouds (Elmegreen et al. 2006; Gouliermis et al. 2014), while other works  have argued that the similarity between the gas distribution and the distribution of stars can be quickly altered by the effects of rapid gas expulsion (Dib 2011; Dib et al. 2011;2013) or by gas tidal shocking of the young star clusters by surrounding gas clouds (Kruijssen et al. 2012). The question of whether the most massive stars in young clusters are mass segregated with respect to the total population of stars is still highly debated and has received a significant amount of attention both in observational studies (Hillenbrand \& Hartmann 1998; de Grijs et al. 2002; Gouliermis et al. 2004; Kerber \& Santiago 2006; Chen et al. 2007; Bontemps et al. 2010a; Er et al. 2013; Habibi et al. 2013; Lim et al. 2013; Elmegreen et al. 2014; Yu et al. 2017; Kuhn et al. 2017; Dib et al. 2018; Moser et al. 2019) and in theoretical/numerical works (Bonnell et al. 2003; Dib et al. 2007a; Dib 2007; Dib et al. 2008b; K\"{u}pper et al. 2011; Olczak et al. 2011; Maschberger \& Clarke 2011; Geller et al. 2013; Sills et al. 2018). The debate expands on whether the observed levels of mass segregation are primordial or due to dynamical interaction between stars in the clusters (Khalisi et al. 2007; Dib et al. 2018).

In this work, we measure the spatial distribution of dense cores in the regions of Aquila, Taurus, Corona Australis, and W43 using the \Q\ parameter (Cartwright \& Whitworth 2004). We also quantify the levels of mass segregation in those regions using a measurement of the mass segregation ratio following the methods of Allison et al. (2009) and Olczak et al. (2011). We explore how the structure of these star forming regions and their levels of mass segregation correlate with their star formation activity. The paper is organized as follows: in \S.~\ref{datasets}, we briefly present the data sets that are used in this study. In \S.~\ref{methods}, we recall the basics of the methods used to quantify the structure parameter and mass segregation ratios. In \S.~\ref{results}, we present our results and in \S.~\ref{discussion}, we discuss them in light of previous work. In \S.~\ref{conclusions}, we conclude. 

\section{The data set of star forming regions}\label{datasets}

\begin{figure}
\centering
 \includegraphics[width=0.90\columnwidth]{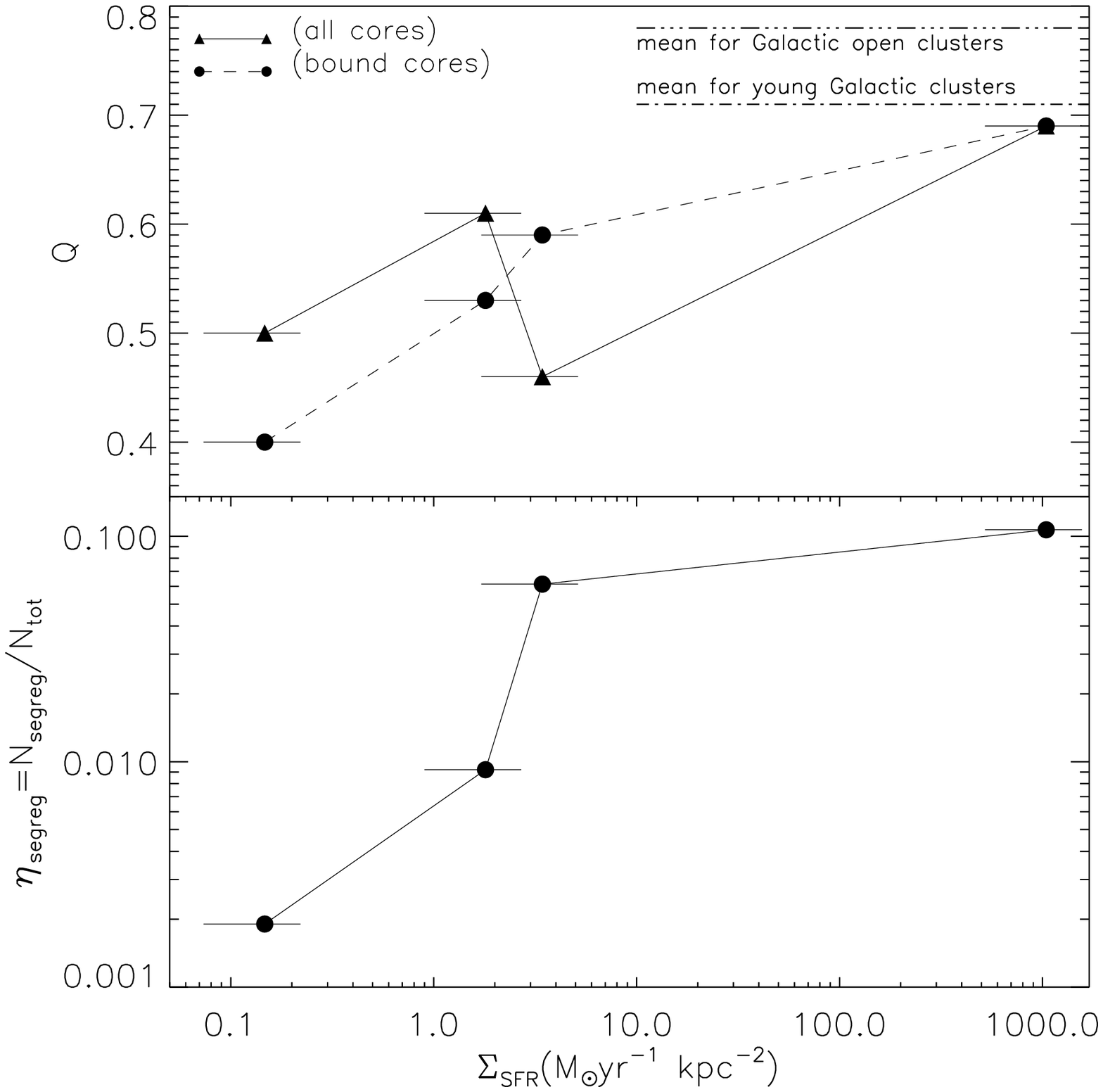}
 \vspace{0.75cm}
 \caption{The correlation between the structure parameter, \Q\, for the entire and bound populations of cores (dashed and full lines, respectively) versus the surface density of star formation $\Sigma_{\rm SFR}$ (top panel) and the correlation between the ratio of the most segregated massive cores to the total number of cores and the $\Sigma_{\rm SFR}$. The horizontal error bars are the $50\%$ uncertainties estimates that are likely to be affecting the measurements of $\Sigma_{\rm SFR}$. The values of \Q\ =$0.78$ and $0.71$ are the ones measured by Dib et al. (2018) for the populations of open clusters and young clusters in the Milky Way, respectively. The estimate of $\Sigma_{\rm SFR}$ for W43 is made using a different approach than for the other three regions (see text for details). The lines connecting the different sets of points are merely shown to guide the eye.}
 \label{fig3}
\end{figure}

We use the publicly available data of the following star forming regions: Aquila, the L1495 region in Taurus, Corona Australis, and the W43-MM1 region. Observations for the first three regions were performed in the framework of the Herschel Gould Belt Survey (And\'{e} et al. 2010) with a wavelength coverage in the submm bands going from 70 to 500 $\mu$m. For all regions, the cores were extracted using the {\it getsources} algorithm (Men'shchikov et al. 2012). In Aquila, the area of the field covered by the Herschel observations is $\approx 11$ deg$^{2}$ and the source extraction permitted the identification of 651 cores of which 446 are gravitationally bound and can be considered as prestellar cores (Andr\'{e} et al. 2010; Bontemps et al. 2010b; K\"{o}nyves et al. 2010,2015). In the cloud Corona Australis, the field covered by the observations has a surface area of $29$ deg$^{2}$. When applied to the Herschel map of the region, {\it getsources} identified 163 cores, out of which 99 are prestellar core candidates (Bresnahan et al. 2018). in Taurus, we use the data related to the identification of dense cores around the L1495 sub-region which covers a surface area of $\approx 8$ deg$^{2}$ (Marsh et al. 2016). The application of {\it getsources} permitted the identification of 525 starless cores and 52 prestellar gravitationally bound cores. The dense cores in the W43 massive star forming region were extracted using {\it getsources} from the 1.3 mm maps using ALMA observations of the W43-MM1 region (Motte et al. 2018). For this region, {\it getsources} returned 131 cores. However, the observations do not distinguish yet whether these cores are unbound or whether they are prestellar in nature. As pointed out in Motte et al. (2018), the characterization of the gravitational boundedness of these cores requires more scrutiny\footnote{Since W43 is located at a much greater distance than the other 3 regions i.e., at a distance of $\approx 5.5$ kpc (Zhang et al. 2014), it is possible that the dense cores in W43 may be fragmented down to scales that are not yet resolved by the ALMA observations. The current spatial resolution in these observations is $\approx 0.01$ pc and should normally be sufficient to resolve low mass cores. However, it is possible that the low mass cores in W43 may be more confined and have smaller sizes in the higher density and pressure environment of W43 than those found in nearby star forming regions. Another aspect is whether the filtering of large scale emission by the interferometric observations can remove a certain population of cores. The ALMA baselines used for the observations in the W43 range from 13 m to 1045 m, which makes it sensitive, at the distance of W43, to emission on all scales of from  $\approx 9.6\times 10^{-3}$ pc and $\approx 0.31$ pc (Motte et al. 2018). The upper limit of these scales is large enough to encompass the entire region. One can also notice that cores are still detected in an evenly distributed way and are not restricted to the higher emission regions that are located in the central region and the 2 filamentary structures.}. For all three regions observed with Herschel, and given the multi-wavelength coverage of the Herschel instruments, the masses of cores are derived using an estimate of the dust temperature that is obtained by fitting the corresponding spectral energy distribution. In the case of W43, the mass estimates of cores obtained by Motte et al. (2018) were derived by combining all available continuum maps of the region and applying the PPMAP software (Marsh et al. 2015) in order to infer a surface density-weighted dust temperature. In general it is worth noting that for all regions considered here, the characterization of the gravitational boundedness of cores was performed using continuum emission data only. A determination of the temperature of the gas in the cores, and measurements of its dynamics, in and around the cores, are crucial in order to better assess their true degree of gravitational boundedness (e.g., Dib et al. 2007b).
       
\section{Methods}\label{methods}

We quantify the spatial distribution and mass segregation of dense cores using the same method as the one presented in Dib et al. (2018). Namely, we use two methods based on a Minimum Spanning Tree (MST) which is the unique set of straight lines ('edges') connecting a given set of points without closed loops, such that the sum of all edge lengths is a minimum (Kruskal 1956; Prim 1957; Gower \& Ross 1969; Schmeja 2011). 

\subsection{The \Q\ parameter}\label{qparam}

We characterize the spatial distribution of the cores with the \Q\ parameter (Cartwright \& Whitworth 2004) which is given by:
 
\begin{equation}
\Q = \frac{\bar{\ell}_{\rm MST}}{\bar{s}}.
\label{eq1}
\end{equation}

We recall that the \Q\ parameter combines the normalized correlation length $\bar{s}$ (i.e., the mean separation length between all cores) and the normalized\footnote{In this work, we follow Schmeja \& Klessen (2006) when defining both normalization terms for $\bar{\ell}_{\rm MST}$ and $\bar{s}$. For $\bar{\ell}_{\rm MST}$, the normalization term is given by $\sqrt{A_{CH}/N}$ (Marcelpoil 1993) where $A_{CH}$ is the area of the convex hull encompassing all cores, and $N$ is the number of cores. The term $\bar{s}$ is normalized by a radius of a circle, $R_{CH}$ whose area is equal to $A_{CH}$. Other authors use different normalizations. Cartwright \& Whitworth (2004) and Parker (2018) normalize $\bar{\ell}_{\rm MST}$ by $\sqrt{A~N}/(N-1)$, where $A$ is the area of a circle whose radius, $R$, is the distance from the furthest point from the structure centre. These authors normalize $\bar{s}$ by dividing it by $R$. Kirk et al. (2016) also use the convex hull area to normalize $\bar{\ell}_{\rm MST}$, but they use the distance between the centre of the convex hull and the most distant point from this centre, $R_{CH,max}$ to normalize $\bar{s}$. More details on the normalization adopted in this work can be found in Appendix B in Schmeja \& Klessen (2006). A discussion of the effects of the different normalizations is presented in Parker (2018).} mean edge length $\bar{\ell}_{\rm MST}$ derived from the MST. The \Q\ parameter allows us to distinguish between structures with a central density concentration and those that are more hierarchical with a fractal substructure. Large \Q\ values ($> 0.8$) are associated with centrally condensed spatial distributions with radial density profiles of the type $\rho(r) \propto r^{-\alpha}$ (with $\alpha > 0$), while small \Q\ values ( $< 0.8$) indicate spatial distributions with a fractal substructure. \Q\ is correlated with $\alpha$ for \Q\ $>0.8$ and anticorrelated with the fractal dimension $D$ for \Q\ $< 0.8$ (Cartwright \& Whitworth 2004, in particular see Figure 5 in their paper). A detailed description of the method, and in particular its implementation used in this study, is given in Schmeja \& Klessen (2006). The \Q\ parameter has been previously used to study the spatial distribution of dense cores and protostars in star forming regions (Guthermuth et al. 2009; Broekhoven-Fiene et al. 2014; Alfaro \& Rom\'{a}n-Z\'{u}\~{n}iga 2018; Parker 2018) and of stars in both young (Cartwright \& Whitworth 2004; Schmeja \& Klessen 2006; Schmeja et al. 2008) and evolved clusters (Gouliermis et al. 2012; Fernandes et al. 2012; Delgado et al. 2013; Kumar et al. 2014; Gregorio-Hetem et al. 2015; Dib et al. 2018). 

\subsection{Mass segregation ratio}\label{msrparam}

Allison et al. (2009) introduced the mass segregation ratio (MSR), $\Lambda_{\rm MSR}$, as a measure to identify and quantify mass segregation in clusters. The method is based on the calculation of the length of the MST, $l_{\rm MST}$, which measures the compactness of a given sample of vertices in the MST. The mass segregation of a cluster is measured by comparing the value of $l_{\rm MST}$ of the $n_{\rm MST}$ most massive stars, $l^{mp}_{\rm MST}$, with the average $l_{\rm MST}$ of k sets of $n_{\rm MST}$ random stars, $\left< l^{\rm rand}_{\rm MST}\right>$. The value of $\Lambda_{\rm MSR}$ is then given by:

\begin{equation}
\Lambda_{\rm MSR}= \frac{\left<l^{\rm rand}_{\rm MST}\right>}{l^{\rm mp}_{\rm MST}}. 
\label{eq2}
 \end{equation}
  
The method has been modified by Olczak et al. (2011) by using the geometric mean rather than the arithmetic mean in order to minimize the influence of outliers. This method works by constructing the MST for the $n_{\rm MST}$ most massive stars and determining the mean edge length $\gamma_{\rm mp}$. Then, we construct the MST of the same number of randomly selected stars from the entire sample and determine the mean edge length $\gamma_{\rm rand}$. The value of the MSR following Olczak et al. (2011), $\Gamma_{\rm MSR}$, is then given by:
 
\begin{equation}
\Gamma_{\rm MSR} = \frac{\langle \gamma^{\rm rand}_{\rm MST} \rangle}{\gamma^{\rm mp}_{\rm MST}}.
\label{eq3}
\end{equation}

\begin{table*}
	\centering
	\caption{Structure parameters and mass segregation levels for the studied star forming regions. The columns refer to: (1) name of the star forming region (2) surface density of star formation (3)-(5) \Q\ parameter, $\Lambda^{max}_{\rm MSR}$, and $\Gamma^{max}_{\rm MSR}$ for the total core population (i.e., label all), and (6)-(8) \Q\ parameter, $\Lambda^{max}_{\rm MSR}$, and $\Gamma^{max}_{\rm MSR}$ for the bound cores population (i.e., label bound).}
	\begin{tabular}{lcccccccc} 
		\hline
	
	           Region & $\Sigma_{\rm SFR}$  &  \Q\ & $\Lambda^{max}_{\rm MSR} (n_{\rm MST})$  & $\Gamma^{max}_{\rm MSR}(n_{\rm MST})$ & \Q\ & $\Lambda^{max}_{\rm MSR} (n_{\rm MST})$  & $\Gamma^{max}_{\rm MSR} (n_{\rm MST})$ \\	
	                           & (M$_{\odot}$ Myrs$^{-1}$ pc$^{-2}$) & (all) & (all) & (all) & (bound) & (bound) & (bound)      \\                
	           \hline
	Taurus\tablefoottext{a}         &    0.14     &   0.50    &     no segregation      &  no segregation   &  0.40  &  no segregation     & no segregation \\
	Aquila                  &    1.8     &   0.61    & 3.75 (6)\tablefoottext{c}                      & 2.53 (6)              &  0.53   & 3.47 (6)                  &  2.36 (6)           \\
	Corona Australis  &     3.43    &   0.46    & 8.80 (10)                  & 12.18 (10)          &   0.59  &  5.68 (11)               & 6.15 (10)          \\
        W43                    &      1043       &   0.69    & 3.49 (14)                   & 2.74 (14)             &   0.69\tablefoottext{b}  &  3.49\tablefoottext{b} (14)  &  2.74\tablefoottext{b} (14)   \\   
		\hline 
	\end{tabular}
	\tablefoot{
	\tablefoottext{a} {Only the region L1495 of Taurus is considered in this work}
	\tablefoottext{b} {This is only valid under the assumption that all cores detected in W43 are gravitationally bound.}
	\tablefoottext{c} {The numbers between parentheses represent the number of most massive cores that are the most mass segregated; i.e., positions at which there is a peak in the $\Lambda_{\rm MSR}$ (or $\Gamma_{\rm MSR}$)-n$_{\rm MST}$ relation.}
	}
	\label{table1}
\end{table*}
 
In this work, we compute both $\Lambda_{\rm MSR}$ and $\Gamma_{\rm MSR}$. In each case, this is done 100 times in order to obtain the mean quantities $\langle l^{\rm rand}_{\rm MST} \rangle$ and $\langle \gamma^{\rm rand}_{MST} \rangle$. A value of $\Lambda_{\rm MSR} \approx 1$ (respectively $\Gamma_{\rm MSR} \approx 1$) implies that both samples of stars (i.e., the most massive and the randomly selected) are distributed in a similar manner, whereas $\Lambda_{\rm MSR} \gg 1$ (respectively $\Gamma_{\rm MSR} \gg 1$) indicates mass segregation, and $\Lambda_{\rm MSR} \ll 1$ (respectively $\Gamma_{\rm MSR} \ll 1$) points to inverse mass segregation, i.e. the massive stars are more spread outwards than the rest.

\section{Results}\label{results} 

We derived the values of \Q, $\Lambda_{\rm MSR}$, and $\Gamma_{\rm MSR}$ for the regions L1495 in Taurus, Aquila, Corona Australis, and W43. For each region, we measure these quantities both for the total population of cores and for the sub-sample of gravitationally bound cores. The latter category includes cores that have been assessed as being prestellar (i.e., gravitationally bound) and cores that already harbor a protostar. The calculations for $\Lambda_{\rm MSR}$ and $\Gamma_{\rm MSR}$ are performed for values of $n_{\rm MST}$ ranging from $n_{\rm MST}=3$ to $n_{\rm MST}=100$. The values we found for these different regions are summarized in Tab.~\ref{table1}. For $\Lambda_{\rm MSR}$, and $\Gamma_{\rm MSR}$ we report in Tab.~\ref{table1} their maximum values ($\Lambda_{\rm MSR}^{max}$ and $\Gamma_{\rm MSR}^{max}$, only if larger than 2) and the corresponding value of $n_{\rm MST}$ at the location of the peak (given between brackets). 

The first observation that can be made is that the \Q\ values of all regions are well below the transition value of $0.8$, which qualifies all of these regions as having a hierarchical structure. These numbers quantify what is already visible by eye in Fig.~\ref{fig1}. An interesting aspect is that the region with the smallest star formation activity, Taurus, displays the smallest value of the \Q\ parameter, particularly for its population of gravitationally bound cores. The higher level of fractal structure in Taurus could be associated with a younger age of the region and/or with processes that are preserving the fractal structure of the cloud for a more extended period of the cloud's lifetime. One of these physical agents could be stronger magnetic fields that are prevalent in Taurus (Chapman et al. 2011; Heyer \& Brunt 2012). The \Q\ parameters for the population of bound cores in these regions display the same behavior, with the bound cores in Taurus exhibiting stronger levels of substructure. The values of \Q\ derived here for dense cores are markedly smaller than those found for open clusters including young clusters (with ages $\lesssim 10-12$ Myrs) which have \Q\  values that lie in the range 0.65-0.8 (Dib et al. 2018). 

In terms of mass segregation, the regions of Aquila, Corona Australis, and W43 show significant levels of mass segregation for the ensemble of the 6, 10, and 14 most massive cores, respectively (see Fig.~\ref{fig2} and Tab.~\ref{table1}). This corresponds to cores with masses in the range [6.2-19.7] M$_{\odot}$, [0.5-1.3] M$_{\odot}$, and [16-102] M$_\odot$, respectively. This is observed both in the $\Lambda_{\rm MSR}$ and $\Gamma_{\rm MSR}$ parameters. In contrast, the cores in Taurus do not show significant signs of mass segregation, within the uncertainties (reasoning in terms of $\Gamma_{\rm MSR}$ instead of $\Lambda_{\rm MSR}$ would imply very mild mass segregation in Taurus, within the $3-\sigma$ uncertainty levels.)\footnote{Fig.~\ref{fig2} is reproduced in Fig.~\ref{figappb1} in App.~\ref{appb} for each star forming region, individually, including the error bars on the measurements of $\Lambda_{\rm MSR}$ and $\Gamma_{\rm MSR}$}. With the exception of Taurus, these results imply that mass segregation is primordial i.e., massive cores, and possibly massive stars (given that current observations show that massive cores display little evidence of fragmentation; e.g., Bontemps et al. 2010), tend to form in the denser parts of the star forming regions rather than fall into them by dynamical effects. The physical mechanism by which this occurs is likely related to the assembly process of the clouds and to the formation of a deeper gravitational potential in these dense regions. In a deep potential well, the accretion of gas by the cores and their mass growth is enhanced, leading to the observed levels of mass segregation (Dib et al. 2010). In contrast, in a region like Taurus with a shallower density profile, both the accretion and the relative motions of cores might be smaller due to magnetic support, and the effects of mass segregation are absent or delayed.  A similar conclusion has been made for dense cores observed in Serpens South by Plunkett et al. (2018)\footnote{Instead of using the position of the peak in the $\Lambda_{\rm MSR}-n_{\rm MST}$ figure to measure what are the cores that are the most mass segregated, Plunkett et al. (2018) look at all $n_{MST}$ cores for which $\Lambda_{\rm MSR} \gtrsim 1$ (within the error bars) to asses the relevance of mass segregation.}.  

In order to check the results of the MST method, we also performed an analysis of the mass segregation levels in those four star forming region using a different approach. The method is based on evaluating whether more massive cores reside in regions of higher surface density of cores, similar to what has been presented in Maschberger \& Clarke (2011), Kirk et al. (2016) and Lane et al. (2016). While the two methods do not reflect the same meaning about what mass segregation is, the findings from this method (i.e., the $\Sigma_{6}-M_{core}$) lend support to our findings using the MST-based approach (see App.~\ref{appa} for more details and discussion). Estimating the effects of mass segregation could be affected by issues related to incompleteness and crowding as discussed by Ascenco et al. (2009) for the case of massive clusters. For all regions that are part of the Gould Belt survey and in observations of W43-MM1, the incompleteness limit is estimated by injecting synthetic sources that are randomly distributed across the high column density parts of the regions (see K\"{o}nyves et al. 2015; Motte et al. 2018). When the CMFs of these different regions are compared for core masses above the $90\%$ completeness limit, differences between those regions persist (i.e., for example in the CMF). It remains to be seen how this picture can be affected when synthetic cores that are injected in the maps are assigned specific spatial distributions and mass segregation levels, and whether these spatial distributions and mass segregation levels can also be recovered. Our intuition is that this would play only a minor effect, since the completeness limit in those regions (apart from W43) is well below the turnover point in each of the CMFs.

We explore whether the value of the \Q\ parameter, both for the total population of cores and for the gravitationally bound ones is related to the star formation activity of the clouds, represented by the surface density of star formation, $\Sigma_{\rm SFR}$. For Taurus, the value of $\Sigma_{\rm SFR}$ was derived using YSO counts from Rebull et al. (2010) and an $A_{V} > 2$ gas mass from Pineda et al. (2010). This yields a value of $\approx 0.14$ M$_{\odot}$ yr$^{-1}$ kpc$^{-2}$ (Heidermann et al. 2010). For Corona Australis, Heidermann et al. (2010) derived a value of $\Sigma_{\rm SFR}=3.43$ M$_{\odot}$ yr$^{-1}$ kpc$^{-2}$ and for Aquila, Evans et al. (2014) obtained a value of 1.8 M$_{\odot}$ yr$^{-1}$ kpc$^{-2}$ which is also computed down to an extinction limit of $A_{V}=2$. The uncertainty from these SFR measurement using YSOs counts is dominated by the uncertainty on the ages of the class II phase which has been assumed by Heidermann et al. to be $50\%$ (i.e., $2\pm1$ Myrs). Using data from Louvet al. (2014), we calculate a value of $\Sigma_{\rm SFR}$ in W43-MM1 of $\approx 1043$ M$_{\odot}$ yr$^{-1}$ kpc$^{-2}$ (see Table 3 in Louvet et al. for the Ridge region)\footnote{{The measurement of the SFR in W43 is performed differently than in the other regions, due to the absence of YSO counts in this region. The SFR is assumed to be given by the estimated total stellar mass that is to be found in cores, divided by a typical timescale for individual star formation of $t_{SF}=0.2$ M$_{\odot}$ (assumed as an mean value over all core masses). The stellar mass that forms in each core is taken as being equal to one third of the mass of the core. Louvet et al. (2014) do not provide uncertainties over the SFR estimate in W43 but it is not unreasonable to think that an uncertainty of $50\%$ seems plausible. This accounts for uncertainties in the core-to-star(s) formation efficiency and to uncertainties over the value of $t_{SF}$}.We can apply the same methodology to derive the $\Sigma_{SFR}$ .for Taurus, Aquila, and CrA. The total mass in pre-stellar cores (i.e., bound cores) in those three regions is 31 M$_{\odot}$, 15.1 M$_{\odot}$, and 470 M$_{\odot}$, respectively. Adopting the same core-to (single) star efficiency of 0.3 and dividing by both $t_{SF}$ and the surface areas of each region listed in Heidermann et al., the corresponding values of $\Sigma_{\rm SFR}$ for Taurus, Aquila, and CrA are $\approx$ 0.18, 3.9 and 7 M$_{\odot}$ yr$^{-1}$ kpc$^{-2}$, respectively.}. Fig.~\ref{fig3} (top panel) displays the correlation between the structure parameter, \Q\, for the entire and bound populations of cores (full and dashed lines, respectively) versus $\Sigma_{\rm SFR}$. A correlation is observed in which clouds that are more hierarchical in nature are associated with lower surface densities of the star formation rate, $\Sigma_{\rm SFR}$. The star formation activity is also correlated with the population of the most massive cores in each region that are found to be strongly mass segregated. Fig.~\ref{fig3} (bottom panel) displays the correlation between $\Sigma_{\rm SFR}$ and the ratio of the number of massive cores that are most mass segregated to the total number of cores in the star forming regions $\eta_{\rm segreg}=N_{\rm segreg}/N_{\rm tot}$. A clear correlation is observed by which a higher star formation activity in the cloud is associated with a larger fraction of mass segregated massive cores\footnote{Moser et al. (2019) measured the \Q\ parameter for the sample of 35 massive protostars in the IRDC G028.37+00.07 and obtained a value of 0.66. They also estimated the SFR in this IRDC to be $1.6\times10^{-3}$ M$_{\odot}$ yr$^{-1}$. Adopting a distance of 5 kpc and a surface area of 184 pc$^{2}$ (estimated from data in Simon et al. 2006) yields $\Sigma_{\rm SFR} \approx 8.7$ M$_{\odot}$ yr$^{-1}$ kpc$^{-2}$. This places this region at the top right of Corona Australis in Fig.~\ref{fig3} and in good agreement with the trends observed for the other regions} . 

The data points in Fig.~\ref{fig3} (top and bottom) are only snapshots of the star forming regions at a given time in their evolution. Dib et al. (2018) showed that the \Q\ parameter distribution for Galactic open cluster peaks at $\approx 0.78$ and the one for young Galactic clusters at $\approx 0.71$. Thus, one would expect that a significant fraction of protocluster forming regions to approach this limit.  The correlation between \Q\ and $\Sigma_{\rm SFR}$ can either be the reflection of a time evolution effect (i.e., regions becoming more centrally condensed as they become more gravitationally bound and which is accompanied by a higher SFR) or otherwise represent an imprint of the cloud formation process. One expects that if a star forming region contracts globally under its own self-gravity, it will likely move in Fig.~\ref{fig3} (top) towards higher \Q\ and $\Sigma_{\rm SFR}$ values. However, for regions such as Aquila and Corona Australis, the ascent along this track will be halted before they can reach substantially higher values of \Q\ and $\Sigma_{\rm SFR}$. This is primarily due to the fact that dense cores have finite lifetimes that are shorter than the regions crossing times. For the L1495 region in Taurus, this time-evolution scenario is even more problematic, as it starts from both a low \Q\ value and a low $\Sigma_{\rm SFR}$ and it is likely that Taurus, if it manages to form any cluster(s), these clusters will simply populate the low \Q\ tail in the distribution measured by Dib et al. (2018). Thus, we favor a scenario in which the currently observed \Q\ - $\Sigma_{SFR}$ correlation is an imprint of the cloud assembly process. When gas is expelled from each of the region proto-subclusters, the evolution towards a more centrally condensed structure with higher \Q\ values that are measured for open clusters (Dib et al. 2018) can proceed in the gas-free phase without being accompanied by an increase in the $\Sigma_{\rm SFR}$. Therefore, we attribute the region-to-region variations of \Q\ and $\Lambda_{\rm MSR}$ and the correlation between these quantities and the star formation activity of the clouds to differences in the physical conditions that have been imprinted on them by the large scale environment at the time they started to form. 

It is important to point out that the correlations observed in Fig.~\ref{fig3} are obtained using only 4 star forming regions. A larger body of observational data is needed before we can derive a more robust quantitative dependence between $\Sigma_{\rm SFR}$ and the structure and mass segregation of dense cores in star forming regions. Nonetheless, these results offer a novel way for looking at the interplay between the structure of molecular clouds and their star formation rates. 

\section{Comparison to previous work and discussion}\label{discussion}

A few studies on individual star forming regions by other group corroborate our findings. Alfaro \& Rom\'{a}n-Z\'{u}\~{n}iega (2018) analyzed the structure and mass segregation of dense cores in the quiescent Pipe nebula cloud which is characterized by a very low star formation activity (Lada et al. 2008). While in our case we chose to present an analysis of \Q\ and $\Lambda_{\rm MSR}$ (and $\Gamma_{\rm MSR}$) for the total and bound population of dense cores, Alfaro \& Rom\'{a}n-Z\'{u}\~{n}iega (2018) took a different approach and measured these parameters for cores selected in different bins of mass and gas density. For the Pipe nebula cloud, they found \Q\ values that are smaller for dense cores/peaks selected at either lower masses or lower gas volume densities of the order of $\approx 0.4$, similar to what we have measured for Taurus. They also found that the $\Lambda_{\rm MSR}$ values are relatively insensitive to the choice of the range of these physical parameters. Using data from the James Clerk Maxwell Telescope (JCMT) Gould Belt Survey, Kirk et al. (2016) applied both an MST based method (on the fluxes of cores rather than their masses) and the surface density-core mass method (see App.~\ref{appb}) and found evidence of primordial flux/mass segregation in the three subregions of Orion B (L1622, NGC 2023/2024, and NGC 2068/2071). Lane et al (2016) performed a very similar analysis to Kirk et al. for the entire Orion A cloud and reached similar conclusion. It is noteworthy that Kirk et al. (2016) reported very high values of \Q\ for the three sub-regions in Orion B, namely \Q\ $=1.18, 0.99$ and $0.91$ for L1622, NGC 2023/2024, and NGC 2068/2071, respectively. This is puzzling, given that in all three sub-regions, there is clear visual evidence of substructure. Parker (2018) re-analyzed the data of Kirk et al. (2016). He reported values of \Q\ that are $< 0.8$ ($\Q=0.72$, $0.65$, and $0.71$ for the sub-regions L1622, NGC 2068/2071, and NGC 20223/2024, respectively), and also variations in the values of $\Lambda_{\rm MSR}$ between those sub-regions with one sub-region (NGC 2023/2024) displaying strong levels of mass segregation ($\Lambda_{\rm MSR} \approx 28,$ for the $4$ most massive cores), another one (NGC 2068/2071) mild levels of mass segregation ($\Lambda_{\rm MSR} \approx 2$) and the third region, L1622, showing no mass-segregation. Parker 2018 argued (see Appendix A in his paper) that the large \Q\ values obtained by Kirk et al. (2016) are due to choices these authors made in how they normalize the quantities $\bar{\ell}_{\rm MST}$ and $\bar{s}$ (see discussion above in \S.~\ref{qparam}). We do not attempt to place the results of Alfaro \& Rom\'{a}n-Z\'{u}\~{n}iega (2018) and Parker (2018) on Fig.~\ref{fig3}. This is primarily due to the fact that the core selection for the Pipe and Orion B clouds were performed using different core extraction methods/algorithms (i.e., the CLUMPFIND and FellWalker algorithms, respectively) and a quantitative comparison using inhomogeneous data sets can lead to misleading conclusions. Nonetheless, the results of these two studies are broadly consistent with our findings in that the low star forming Pipe molecular cloud has a small \Q\ parameter, similar to the one we derived for Taurus, while the more intensely star forming regions in the Orion B cloud have a higher \Q\ values. Stutz \& Gould (2016) showed that dense cores in Orion A are well connected, both spatially and dynamically, to the integral shaped filament and the underlying gas structure which sets the gravitational potential of the region. Their approach is also another way of looking at the issue of whether there is a connection between the morphology of a star forming region, as traced by its population of dense cores, and the star formation activity in the region. 
 
The existence of substructure in star forming regions is a direct consequence of turbulent fragmentation. When turbulence is injected into the cloud on large physical scales (i.e., scales equal or larger than than cloud scale), a natural consequence of the turbulent cascade is the formation of a network of compressed, post-shock regions of different sizes (Dib et al. 2007b, Federrath et al. 2010; Burkhart et al. 2012; Padoan et al. 2014). Furthermore, as these substructures continue to fragment to smaller scales, forming dense cores, a non-zero level of mass-segregation can be expected, as more massive cores are statistically more likely to form in more massive substructures due to the availability of a larger mass reservoir (Padoan \& Nordlund 2002). In supersonic clouds that are magnetically subcritical such as Taurus \footnote{Hildebrand et al. 2009 and Chapman et al. 2011 report values of the mass-to magnetic flux ratio of $\mu=0.1\pm0.02$ and $0.32\pm0.02$, respectively, where $\mu$ is expressed in units of the critical value for collapse (Nakano \& Nakamura 1978). In W43, $\mu$ is highly supercritical and is found to be much in excess of unity across the entire region (Cortes et al. 2016); i.e., in the range $\approx 10-60$), and in Aquila, the available data about the magnetic field strength (Sofue \& Nakanishi 2017) and mean column density (K\"{o}nyves et al. (2015) suggest that $\mu$ falls in the range $5-7$ using the formula $\mu=7.6 \times 10^{-21} (N(H_{2})/B_{los}$ (Troland \& Crutcher 2008). There is no available information about the magnetic field strength in Corona Australis.}, star formation, which is mediated by ambipolar diffusion, proceeds at a slower pace due to effects of magnetic pressure which prevents substructure from merging efficiently (Nakamura \& Li 2005; Dib et al. 2007b). Thus, a higher level of substructure in the clouds could be indicative of either a young age for the region and/or of the existence of a strong magnetic support. 

With regard to the question of whether mass segregation is primordial (i.e., set at the dense core phase) or induced by the dynamical evolution of an initially sub-structured cluster, Dom\'{i}nguez et al. (2017) argued that a young cluster will quickly settle into a state where the most massive stars are mass segregated, regardless of the existence/absence of mass segregation within its different levels of substructure. On the other hand, Dib et al. (2010) argued that mild levels of mass segregation that can be generated by turbulent fragmentation could be significantly enhanced by gas accretion onto the cores. Dib et al. (2007a) proposed that another efficient channel for imprinting a significant level of primordial mass segregation is by the coalescence of cores. The process is more efficient in the densest regions of protocluster clouds where cores are more closely packed, and it may be the dominant mode of star formation in the centre of massive starburst clusters such as Arches and NGC 3603.

\section {Conclusion}\label{conclusions}

We have analyzed the spatial distribution and mass segregation of dense cores in a number of nearby (the L1495 region in Taurus, Corona Australis, and Aquila) and more distant (W43) star forming regions. The observations for these regions were performed using the Herschel space telescope for the first three regions, and the Atacama Large Millimeter Array for W43. Cores were extracted in those four regions with a similar technique using the {\it getsources} algorithm (Men'shichikov et al. 2012). For each region, we quantify the spatial distribution of the cores using the \Q\ parameter and the mass segregation ratios $\Lambda_{\rm MSR}$. Our analysis indicates that the spatial distribution of dense cores in those regions is very substructured. Given the relatively short lifetimes of most of these cores, it is thus very likely that the emerging young clusters in those regions will inherit the same level of substructure that is found for their parent core population. With the exception of Taurus, we observe mild-to-significant levels of mass segregation for the {ensemble of the 6, 10, and 14 most massive} cores in Aquila, Corona Australis, and W43, respectively. 

We show, for the first time, that the spatial distribution of dense cores is positively correlated with the star formation activity in the clouds, represented by the surface density of the star formation rate, $\Sigma_{\rm SFR}$. Regions that have the lowest star formation activity (i.e., Taurus) are those that display higher level of substructure and no signs of mass segregation of the cores. In contrast, regions with an intense star formation activity (W43) have a higher \Q\ value which is indicative of a larger role by gravity in assembling a more centrally condensed structure. We also show that the ratio of the most massive cores that display the strongest levels of mass segregation to the total number of cores is also correlated with the $\Sigma_{\rm SFR}$. Our findings could be affected, to a limited extent, by effects of time evolution, with substructures merging due to gravity and forming a more centrally condensed distribution. However, we argue that are likely due to a dependence on the physical conditions in the clouds and which have been imprinted on them by the large scale environment at the time these clouds started to form. The analysis of a larger sample of clouds will allow us to shed more light on the interplay between star formation in clouds and the spatial distribution of dense cores, and help us better understand how star clusters form and evolve.

\begin{acknowledgements}
S. D. would like to thank Stefan Schmeja for insightful discussions on issues related to the characterization of structures using the minimum spanning tree method. We thank Vera K\"{o}nyves, Thomas Nony, Davide Elia, Mark Heyer, Sylvain Bontemps, and Jouni Kainulainen for useful discussions on the star forming regions studied in this work. We also thank the anonymous referee for useful comments. This research has made use of data from the Herschel Gould Belt survey project and data from the ALMA-IMF project. S. D. acknowledges support from the french ANR and the german DFG through the project "GENESIS" (ANR-16-CE92-0035-01/DFG1591/2-1).

\end{acknowledgements}

%
%

\begin{appendix}

\section{Measuring mass segregation using the surface density-mass plot}\label{appa}

\begin{figure*}
\centering
\includegraphics[width=0.46\textwidth]{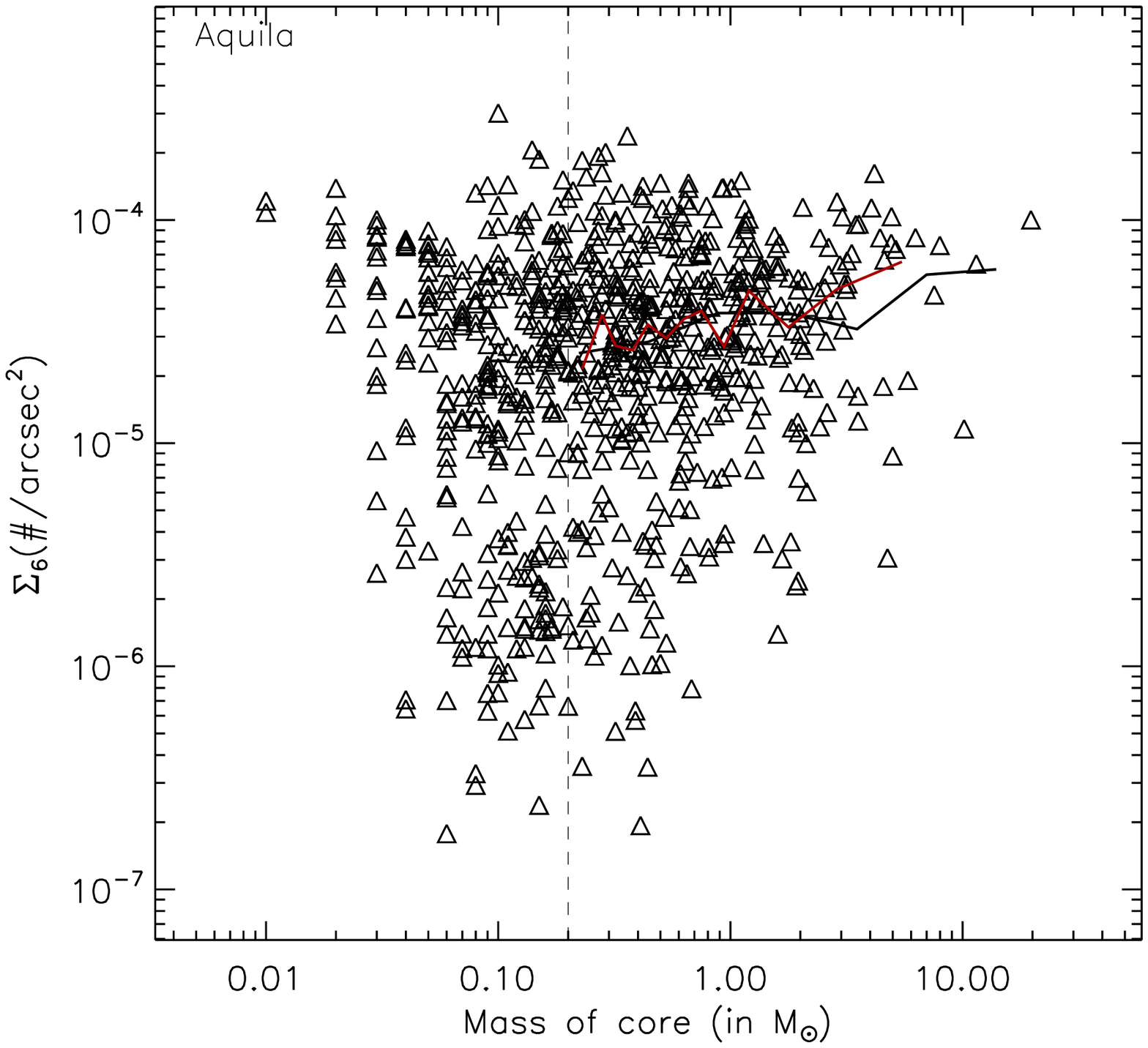}
\includegraphics[width=0.46\textwidth]{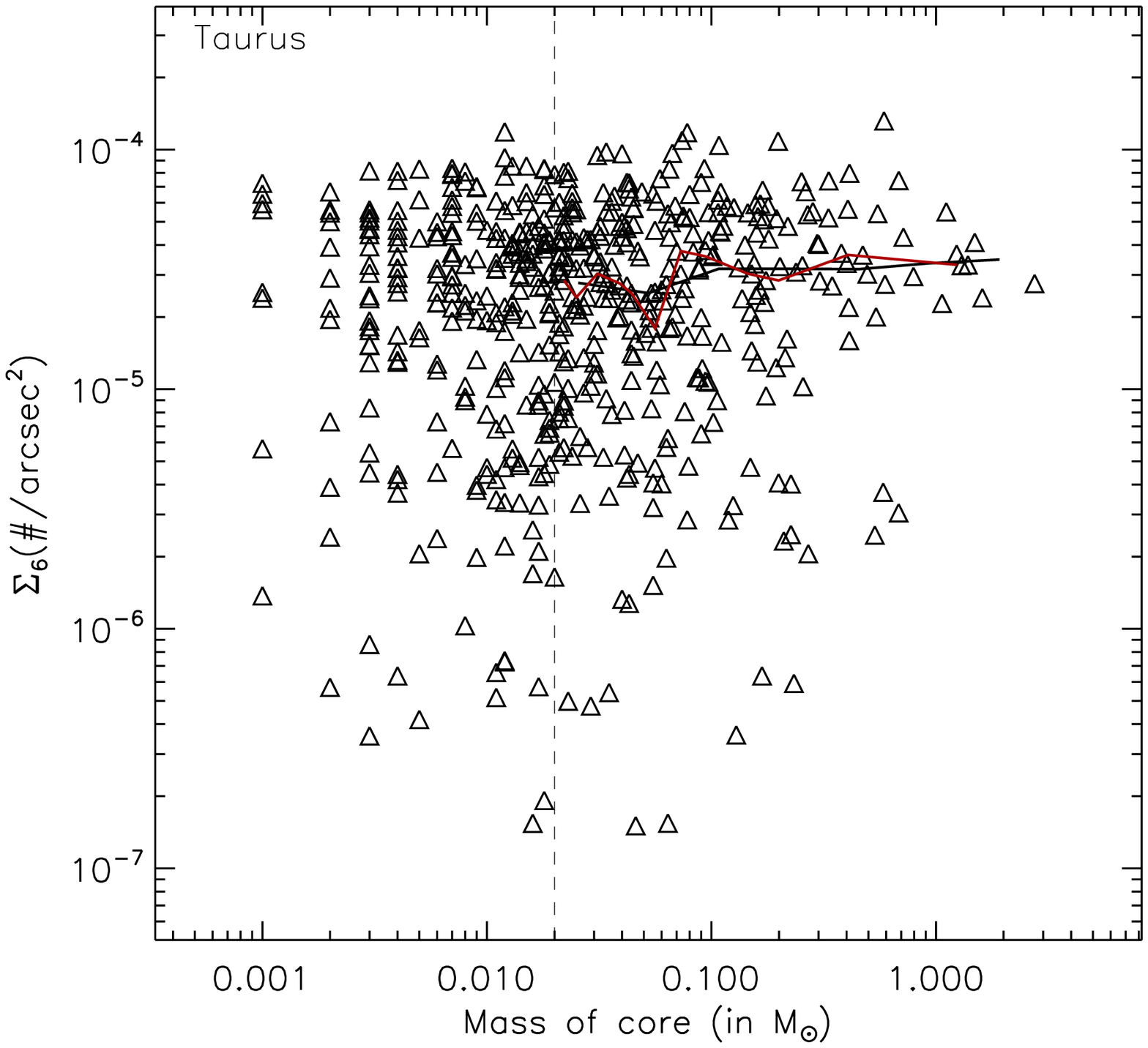}\\
\includegraphics[width=0.46\textwidth]{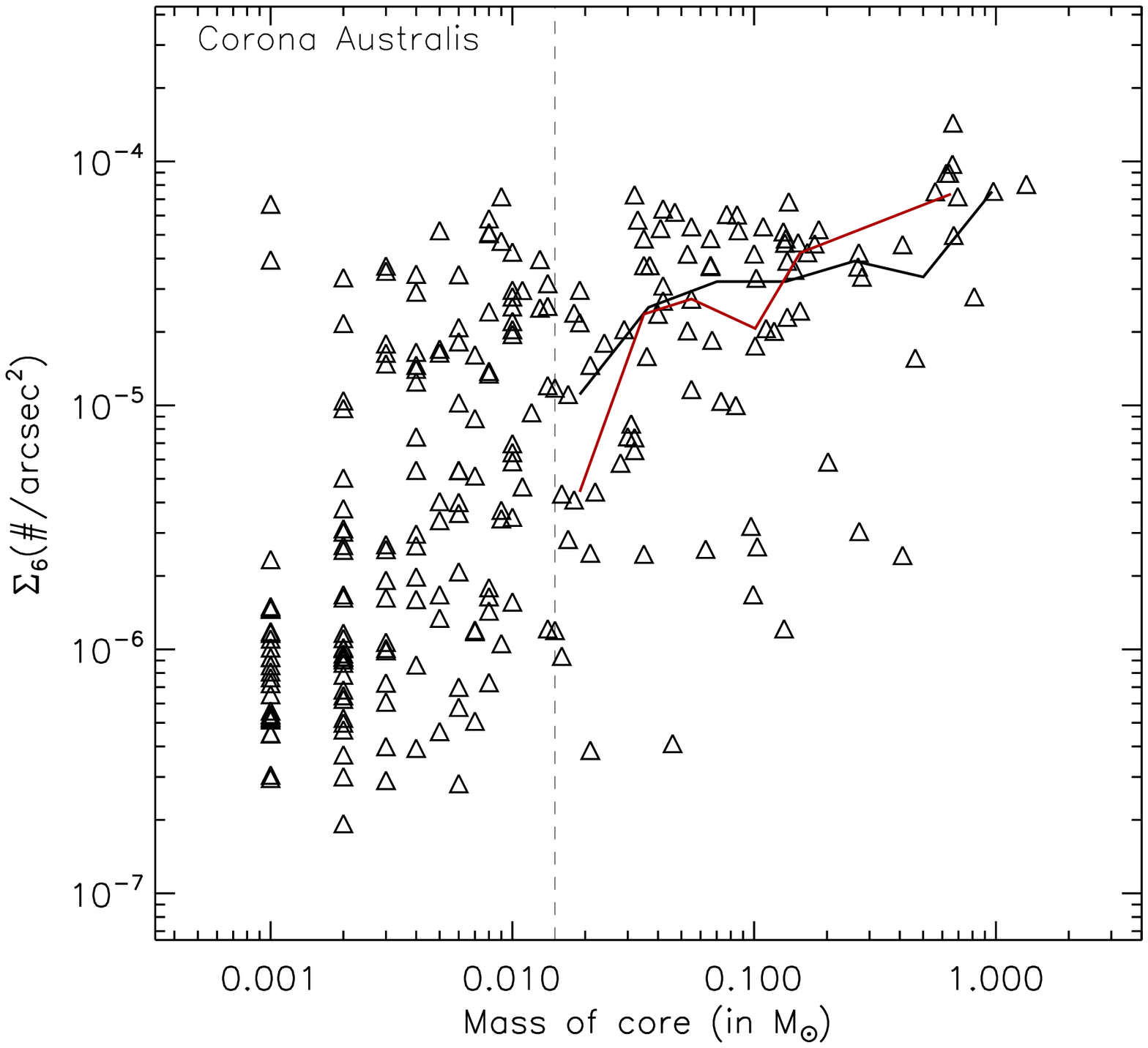}
\includegraphics[width=0.46\textwidth]{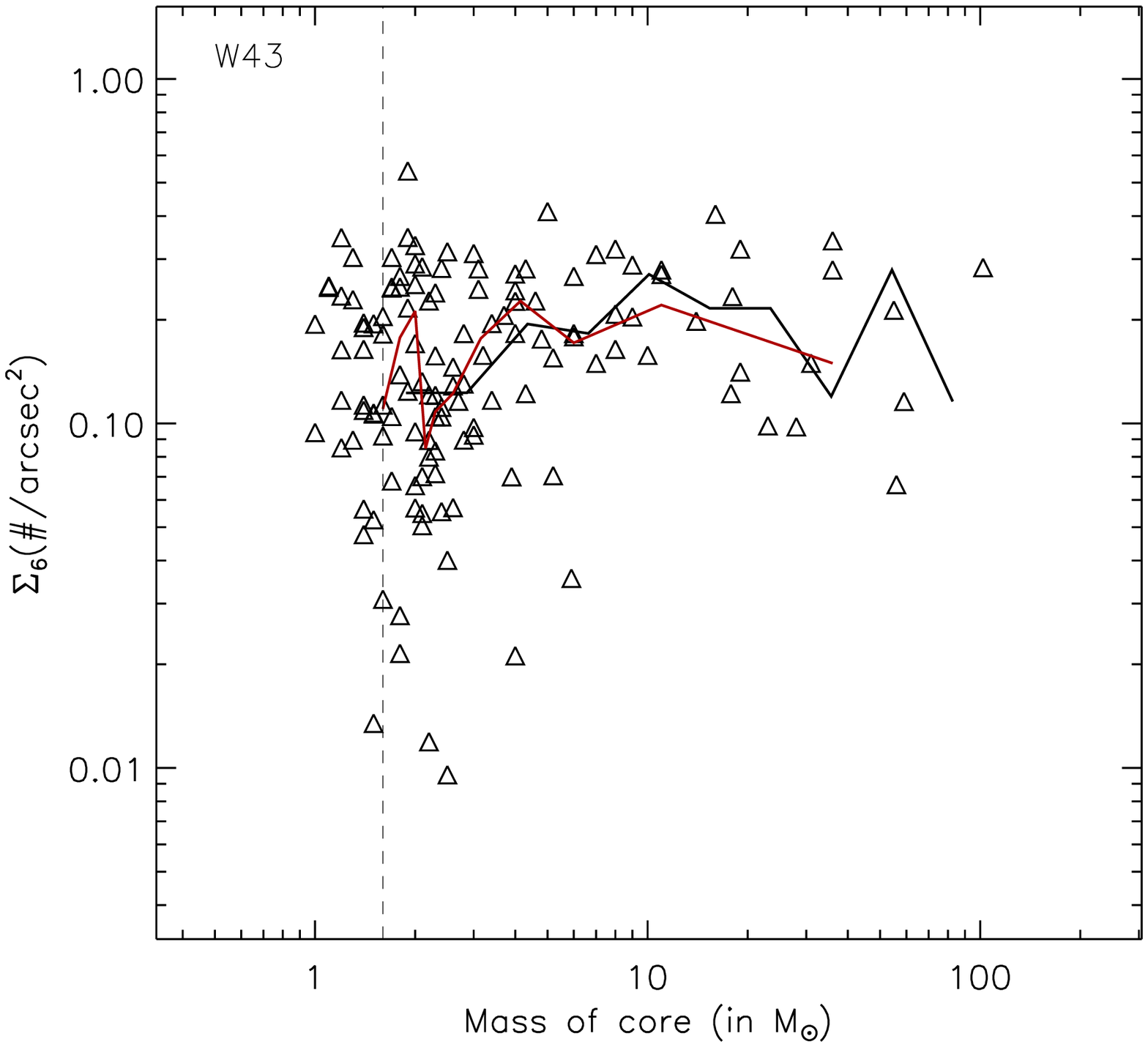}
\caption{Surface density of cores around each core versus its mass. In calculating the surface density around each core, we use the circular area enclosed within the distance to its sixth nearest neighbor. The vertical dashed lines in all subsets show the $90\%$ completeness limit. The full line represents the median value of $\Sigma_{6}$ for bins with a fixed value of log$M$ (black line) and for a variable bin size which contains an equal number of objects (red line).The median values of $\Sigma_{6}$ are shown for values of $M_{core}$ that are above the completeness limit in each region.}
\label{figappa1}
\end{figure*}

Here, we test a different method aimed at characterizing the level of mass segregation in the four star forming regions considered in this work. The method is based on inspecting the relationship between the local surface density of cores around each core and the core mass. Such an approach have been employed in the context of stellar clusters (e.g., Casertano \& Hut 1985; Maschberger \& Clarke 2011) and also in the context of galaxy groups (e.g., Dariush et al. 2016). The local surface density is calculated as being: 

\begin{equation} 
\Sigma_{j}\left({\rm objects/area}\right)=\frac{j-1}{{\rm{\pi}} r_{j}^{2}},
\label{eqappa1}
\end{equation}

where $r_{j}$ is the distance to the $j$-th nearest neighbor to the core. We use a value of $j=6$. Casertano \& Hut (1985) showed that, for a number of discrete objects between 30 and 1000, as in our case, this value is a good compromise between estimating the local densities and the number of density fluctuations in the overall structure\footnote{We varied $j$ between 5 and 8 and saw no qualitative difference in the surface density-mass plot for the different regions.}. Fig.~\ref{figappa1} displays the surface density of cores versus core mass for the four star forming regions considered in this work. The vertical dashed lines in all subsets show the $90\%$ completeness limit. The limits for completeness are taken from K\"{o}nyves et al. (2015) for Aquila, Motte et al. (2018) for W43, and Bresnahan et al. (2018) for Corona Australis. The threshold value for completeness in the L1495 region of Taurus has not been reported in Marsh et al. (2016). However, given that Taurus and Corona Australis are at nearly identical distances ($\approx$ $140$ pc and $\approx 130$ pc, respectively, we adopt a conservative value $0.02$ M$_{\odot}$, only slightly higher than the value of $0.015$ M$_{\odot}$ in Corona Australis. The full line in each subset of the figure shows the median values of $\Sigma_{6}$ calculated in 11 bins of ${\rm log}(M)$, {and the red line} displays the median value calculated in bins that contain the same number of cores. Before discussing the results of the $\Sigma_{6}-M_{core}$ method for the the individual star forming region, it is important to stress that the two method are inherently different. The MST-based method uses information about the positions and masses of all cores, whereas the second method uses only the local position of the j-nearest cores around each core. As such, the $\Sigma_{6}-M_{core}$ can be used to determine whether massive cores reside in higher density environments, but by design, the method would be unable to specify whether massive cores are spatially mass segregated (i.e., if they have a more compact configuration that the rest of the population of cores). Results from the two method could be affected differently by the effect of completeness. Since the completeness limit is higher in higher density environments, more low mass cores could be missed in the direct vicinity of massive cores, thus forcing the method to seek further out the $j$ nearest neighboring cores. The MST-based method would be less affected by this as it uses the positions of a much larger sample of cores that is distributed across the entire region. What can make the  two methods give similar results (but not necessarily with the same level of sharpness) is the hierarchical nature of molecular clouds which is essentially the consequence of turbulent fragmentation. The largest number of objects, including the most massive objects, will statistically be found in the largest structures, so when mass segregation is present, the two methods will indicate that massive cores are more closely packed and that they reside in higher density environments. The self-similar nature of molecular clouds (and of young clusters) can be affected by additional processes such as strong magnetic fields and in the case of young stellar clusters by stellar ejections.  

In the case of the four star forming regions studies in the work, the $\Sigma_{6}-M_{core}$ plots tend to reflect the discussion above. As with the $\rm {MST}$-based method, the $\Sigma_{6}-M_{core}$ plot for Taurus shows that for all core masses, including massive cores, they reside in environments with a (broad) range of surface densities and that there is no correlation between $\Sigma_{6}$ and $M_{core}$. In contrast, in the three other regions, high mass cores reside preferentially in regions of higher surface density. In the Corona Australis region, there is a noticeable increase of $\Sigma_{6}$ with increasing core mass. This is the region that shows the strongest signal in $\Lambda_{\rm MSR}$ (See fig.~\ref{fig2}). For Aquila, the increase in $\Sigma_{6}$ with core mass is milder, but this is still reflected by an increase of $\Sigma_{6}$ with $M_{core}$ by a factor of $\approx 4$. In the case, of W43, the result from the $\Sigma_{6}-M_{core}$ is less conclusive. In this region, all massive cores (i.e., cores with masses $\gtrsim 7$ M$_{sol}$) reside in high density environments, however, there is no clear increase of $\Sigma_{6}$ with $M_{core}$. This could be due to the more significant effects of completeness around the most massive cores, as discussed above. Finally, it is also important to realize that the results of this method, albeit being here in good qualitative agreement with results based on the ${\rm MST}$ method, they do suffer from a poor definition of the surface density of cores. Since the calculation of the area is based on the circular area enclosed within the $j$-th neighbor, the method inherently can become less accurate for massive objects that may be surrounded by many low mass objects but that reside in filamentary structures. A more accurate estimate of the surface area is needed. One possibility is to use the convex hull area, but such details are beyond the scope of this paper and are left for future work.        
 
\section{Uncertainties on the measurement of the mass segregation ratio}\label{appb}

We have omitted showing the uncertainties associated with the measurement of the mass segregation ratios in Fig.~\ref{fig2} for the sake of clarity. While the discussion on how the uncertainties on both $\Lambda_{\rm MSR}$ and $\Gamma_{\rm MSR}$ are measured has been described in detail in Dib et al. (2018 ) and in the original works (Allison et al. 2009; Olczak et al. 2011), we provide here a reminder of the procedure and include the error bars for each of the regions studied in this work, individually. The uncertainty on $\Lambda_{\rm MSR}$ is given by:

\begin{equation}
\Delta \Lambda_{\rm MSR}=\Delta l^{\rm rand}_{\rm MST}
 \label{eqappb1}
 \end{equation}
 
\noindent where $\Delta l^{\rm rand}_{\rm MST}$ is the standard deviation from the $k$ random sets (we use $k=100$ random sets). The standard deviation of $\Gamma_{\rm MSR}$ is given by:

\begin{equation}
\Delta \Gamma_{\rm MSR}=\Delta \gamma^{\rm rand}_{\rm MST}.
\label{eq6}
\end{equation}

\begin{figure*}
\centering
\includegraphics[width=0.45\textwidth]{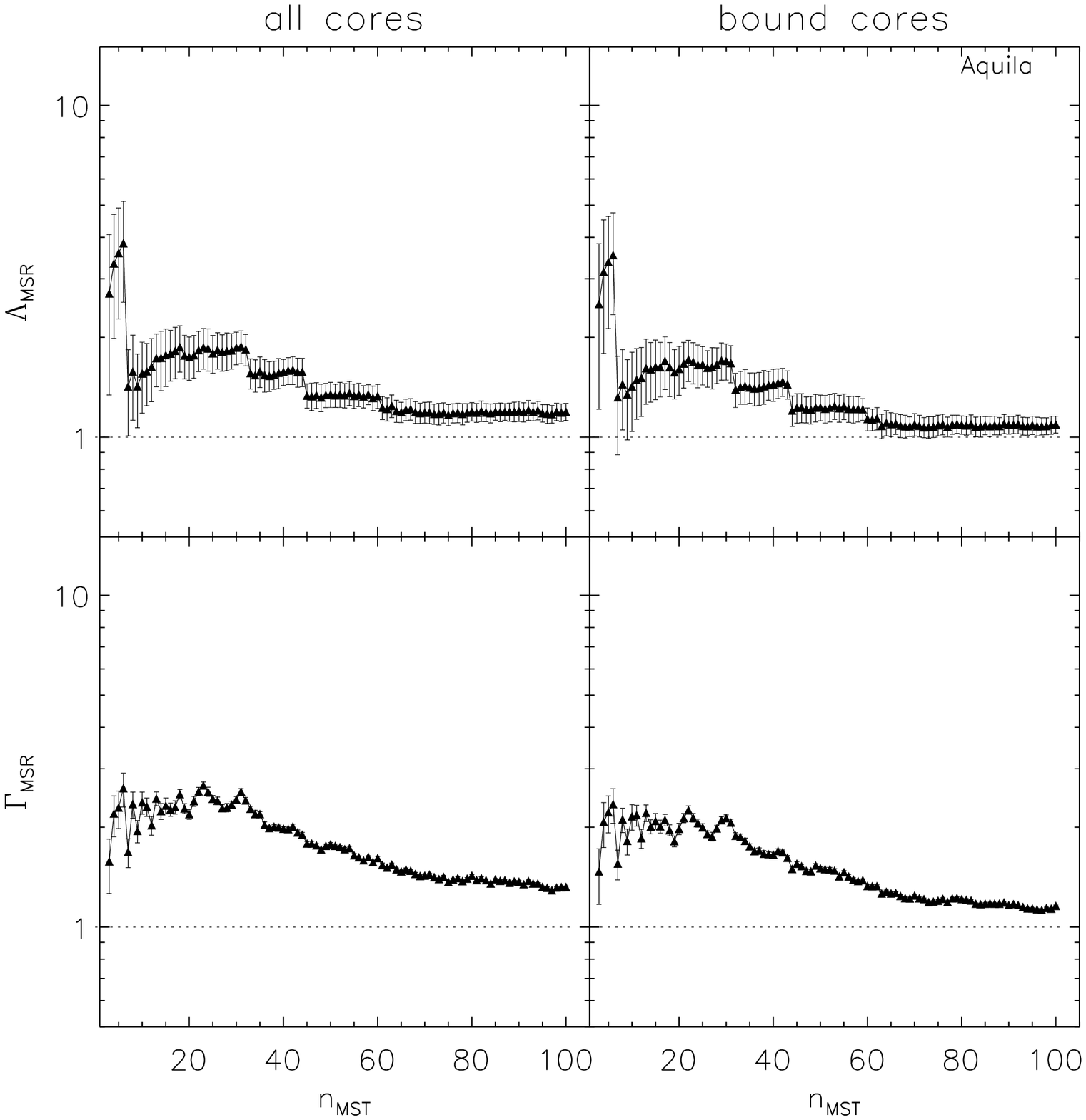}
\vspace{0.75cm}
\hspace{0.75cm}
\includegraphics[width=0.45\textwidth]{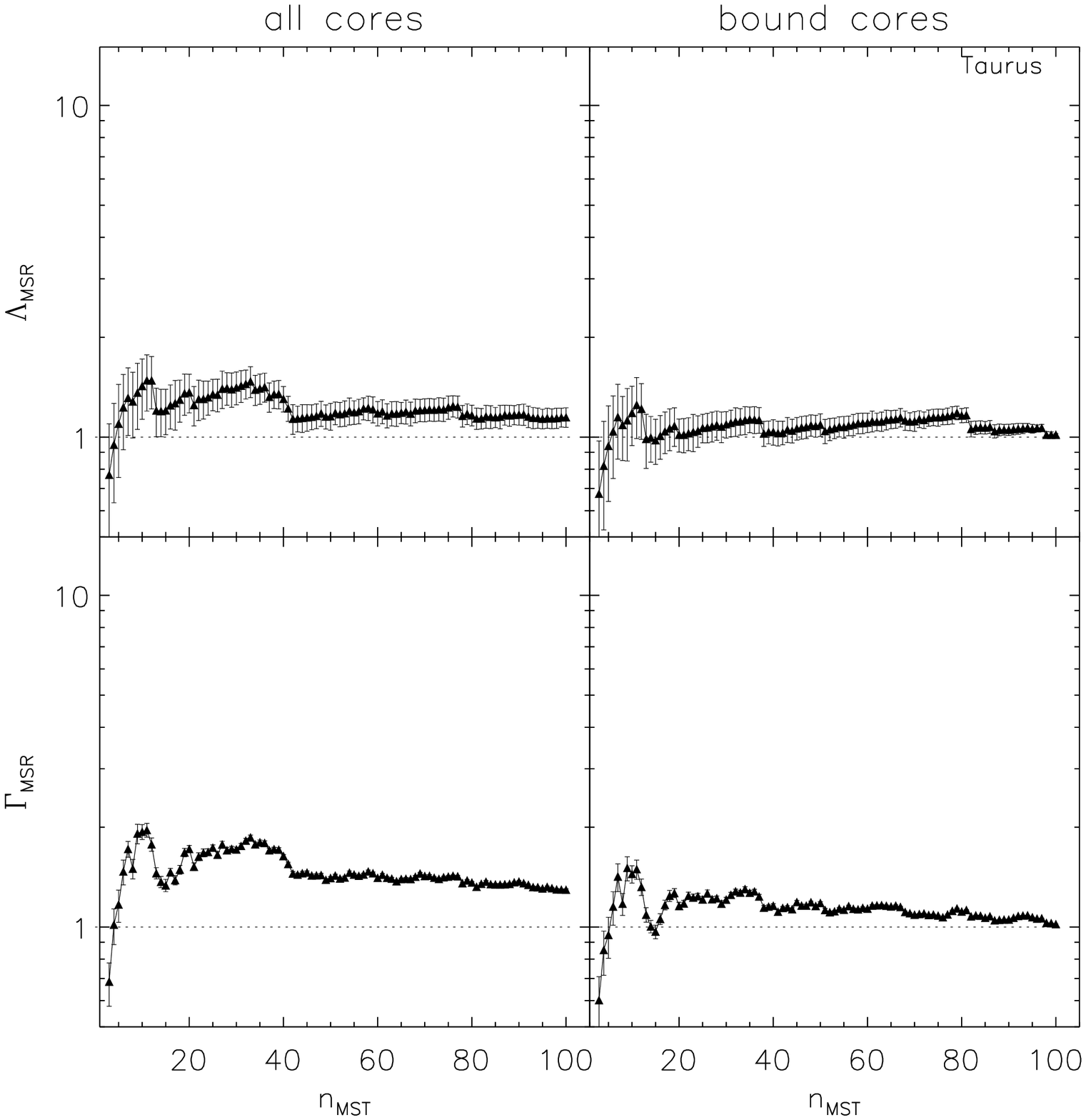}\\
\vspace{0.75cm}
\includegraphics[width=0.45\textwidth]{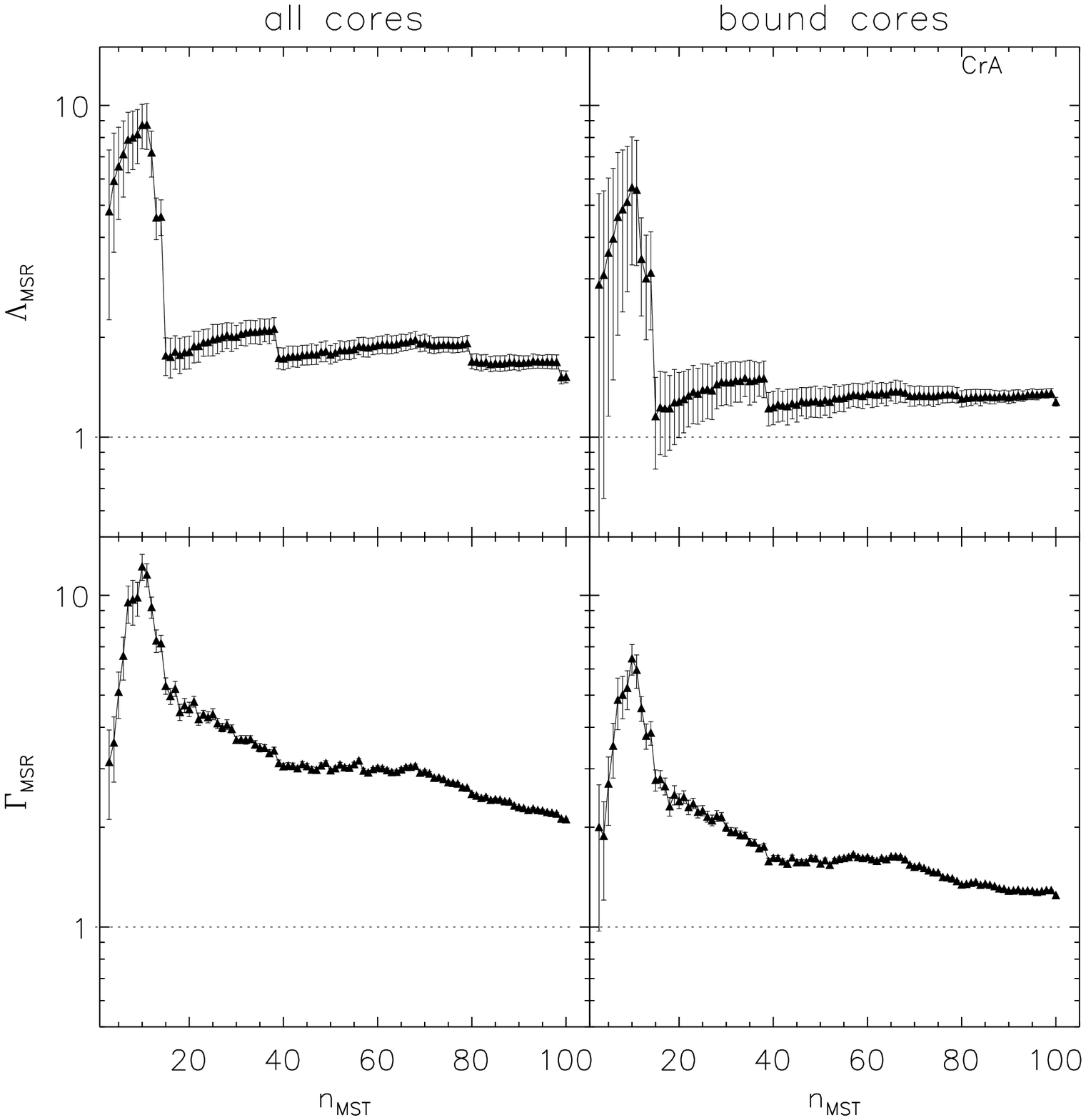}
\hspace{0.75cm}
\includegraphics[width=0.45\textwidth]{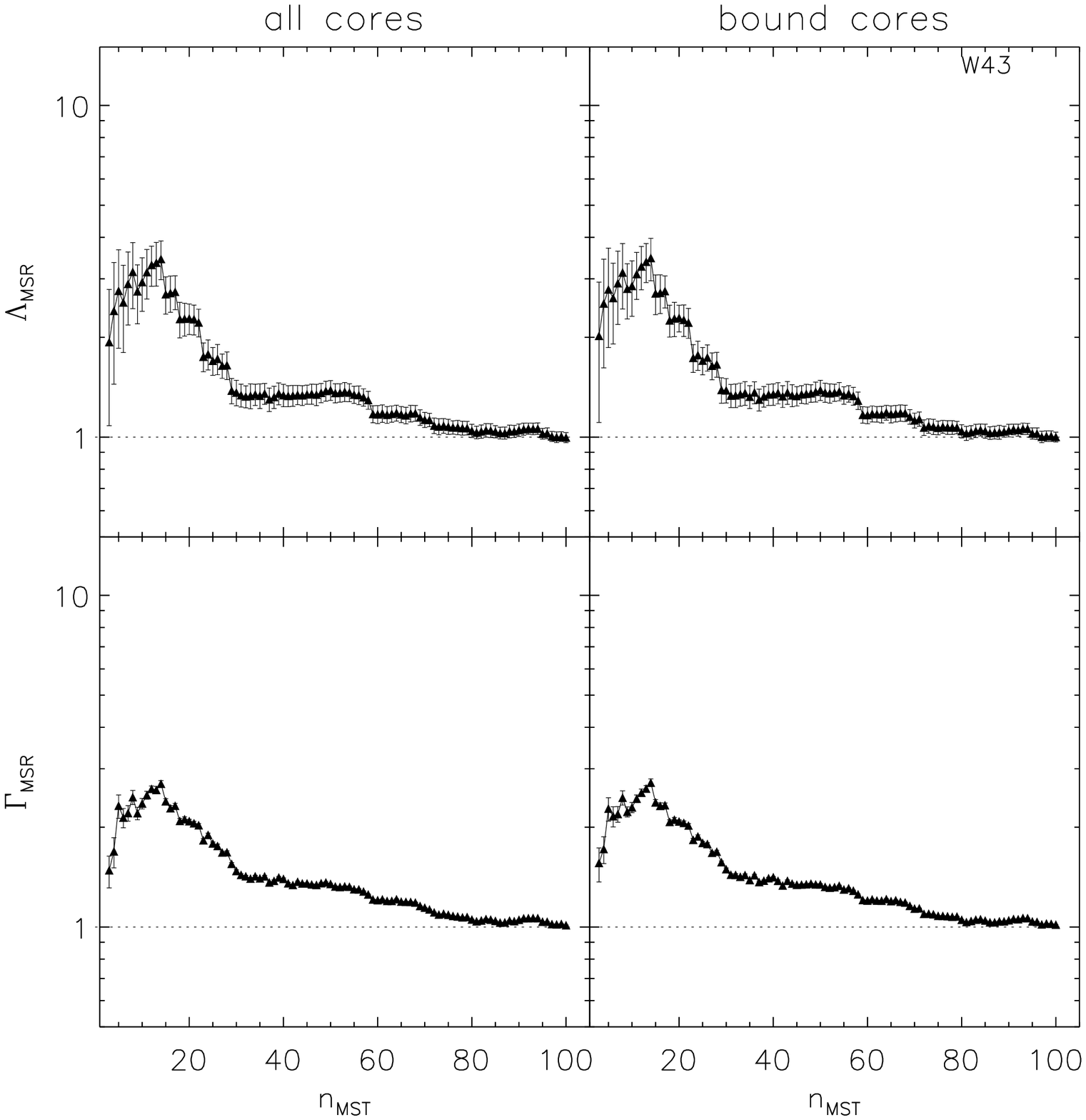}
\vspace{0.75cm}
\caption{Same as Fig.~\ref{fig2} but for the 4 regions shown separately and including the error bars on $\Lambda_{\rm MSR}$ and $\Gamma_{\rm MSR}$.}
\label{figappb1}
\end{figure*}

In Fig,~\ref{figappb1}, we reproduce Fig.~\ref{fig2} for each star forming region, individually and show the associated error bars on $\Lambda_{\rm MSR}$ and $\Gamma_{\rm MSR}$ in the ($\Lambda_{\rm MSR}$,$\Gamma_{\rm MSR})-n_{\rm MST}$ plots. The inclusion of these error bars do not affect our conclusions. 
 
\end{appendix}

\end{document}